\documentclass[%
 aip,
 amsmath,amssymb,
 reprint,%
]{revtex4-1}

\usepackage{graphicx}
\usepackage{dcolumn}
\usepackage{bm}

\usepackage[utf8]{inputenc}
\usepackage[T1]{fontenc}
\usepackage{mathptmx}
\usepackage{etoolbox}
\usepackage{braket}
\newcommand\redsout{\bgroup\markoverwith{\textcolor{red}{\rule[0.5ex]{2pt}{0.4pt}}}\ULon}
\newcommand{\ve}[1]{\mathbf{#1}}
\makeatletter
\def\@email#1#2{%
 \endgroup
 \patchcmd{\titleblock@produce}
  {\frontmatter@RRAPformat}
  {\frontmatter@RRAPformat{\produce@RRAP{*#1\href{mailto:#2}{#2}}}\frontmatter@RRAPformat}
  {}{}
}%
\makeatother
\begin{document}

\preprint{AIP/123-QED}

\title[]{Enhanced quantum correlations from joint pump and photon pair scattering}
\author{M. Safadi\textsuperscript{*}}
\email[Author to whom correspondence should be addressed: ]{mamoon.safadi@mail.huji.ac.il}
\thanks{These authors contributed equally to this work.}
 \affiliation{ 
Racah Institute of Physics, The Hebrew University of Jerusalem, Jerusalem 91904, Israel
}
\author{N. Kuchuk}%
\thanks{These authors contributed equally to this work.}
 \affiliation{ 
Racah Institute of Physics, The Hebrew University of Jerusalem, Jerusalem 91904, Israel
}%
\author{O. Lib}%
 \affiliation{ 
Racah Institute of Physics, The Hebrew University of Jerusalem, Jerusalem 91904, Israel
}%

\author{Y. Bromberg}%
\affiliation{ 
Racah Institute of Physics, The Hebrew University of Jerusalem, Jerusalem 91904, Israel
}%

\author{A. Goetschy}
\affiliation{%
ESPCI Paris, PSL University, CNRS, Institut Langevin, 1 rue Jussieu, F-75005 Paris, France
}%


\begin{abstract}
Scattering of non-classical light is enabling new ways to study and control photon transport. However, advances in this field often rely on simplifying assumptions regarding the quantum light's generation and its source. In this work, we relax some of these assumptions and probe the behavior of entangled photon pairs passing through a disordered layer after being generated by a randomly scattered pump via spontaneous parametric down conversion. We experimentally demonstrate that, even when both the pump and the down-converted photons propagate through a dynamic scattering medium, the pairs maintain a sharp peak in their correlations. A comprehensive theoretical and numerical analysis shows that these correlations persist regardless of when the pairs are generated, whether immediately after the pump is scattered or under other conditions. More specifically, we detail how the shape of the angular correlation depends on the pump’s scattering and how it varies with the distance between the pair-generation region and the entrance of the disordered medium.  These findings represent a crucial step toward understanding quantum light generation in complex media, and potentially exploiting it for quantum technologies.
\end{abstract}

\maketitle

\section{\label{sec:level1} Introduction}

\begin{figure}[hbt!]
 \centering
 \includegraphics[scale=0.65]{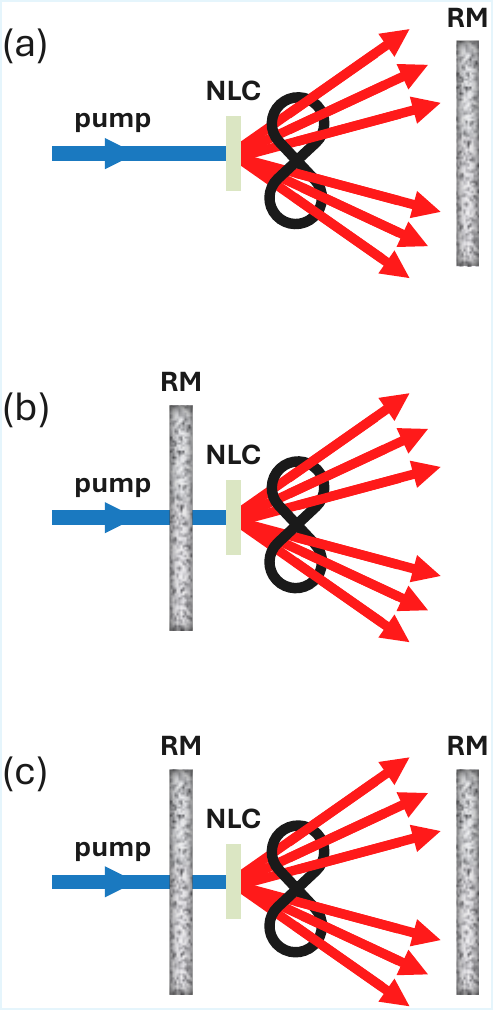}
 \caption{(a) Post-generation scattering: The entangled photon pairs enter a random medium after being generated via spontaneous down conversion (SPDC) from a typical plane-wave pump field. (b) Pre-generation scattering: The entangled pairs are generated from a scattered pump field. (c) Pre- and Post-generation scattering: Both the pump field and the entangled pairs generated are scattered. NLC, nonlinear crystal; RM, random medium.}
\newpage
 \label{fig:1}
\end{figure}

Over the past few decades, scattering of non-classical light has emerged as a rich field of study offering insights and applications that extend well beyond the classical realm \cite{lib2022quantum}. For example, researchers have shown that spatially entangled photon pairs traversing through a static disordered medium [Fig.~\ref{fig:1}(a)] exhibit speckles, coined two-photon speckle, in the angle-resolved two-photon correlations \cite{beenakker2009two,ott2010quantum,cande2013quantum,cande2014transmission,schotland2016scattering}, with entanglement surviving multiple scattering upon probing all output modes \cite{cande2014transmission,valencia2020unscrambling}. Typically, interference effects wash out upon ensemble averaging over realizations of the disorder \cite{pors2011transport,ibrahim2013orbital,krenn2015twisted,leonhard2015universal}; yet, coherent features can still emerge. In a transmission geometry, the two-photon correlation map has revealed an enhancement of the non-classical light that depends on the properties of the entangled state, such as its symmetry \cite{van2012bosonic} and its correlation size \cite{lib2022thermal}. In reflection geometry, we have recently observed the analog of coherent backscattering (CBS) of classical light \cite{Kuga1984,VanAlbada1985,Wolf1985,Akkermans1986} with entangled photon pairs\cite{safadi2023coherent}. This phenomenon, coined two-photon CBS (2p-CBS), manifests as a 2-to-1 enhancement after disorder averaging, with a width determined by the fundamental scattering properties of the sample, primarily its transport mean free path.

When the pump beam that drives the generation of entangled photon pairs in a nonlinear crystal is itself scattered, it opens a new perspective on how disorder and nonlinearity interplay in the generation process [Fig.~\ref{fig:1}(b)]. In particular, wavefront shaping of the pump beam has been shown to control the spatial correlations of entangled pairs and compensate for scattering, allowing for the optimization of the coincidence map using well-established classical tools \cite{lib2020real, lib2020pump, shekel2021shaping, boucher2021engineering}. Moreover, the spatial coherence of the pump not only governs the structure of the generated correlations, but also affects the overall entanglement dimensionality \cite{defienne2019spatially}. Despite these advances, no attention has been given to scenarios in which both the pump and the entangled pairs undergo scattering [Fig.~\ref{fig:1}(c)], even though such conditions naturally arise when pairs are generated inside a nonlinear disordered medium. In contrast, the reversed classical process, second harmonic generation in random media, has been extensively studied, both experimentally and theoretically \cite{agranovich1988effects, makeev2003second, fischer2006broadband, faez2009experimental, valencia2009weak,qiao2019cavity,savo2020broadband,muller2021modeling, morandi2022multiple,samanta2022speckle,samanta2023photon,moon2023measuring, samanta2025second, nardi2025mesoscopic}.
Understanding the joint scattering of the pump and entangled pairs is therefore essential for uncovering the principles that enable coherent pair generation in complex environments.

In this work, we theoretically and experimentally study the properties of entangled pairs when both the pump and the entangled pairs are scattered by a dynamic random medium. We discover that despite the pump being scattered and scrambled, the down-converted photons, which also scatter, retain enhanced correlations after disorder averaging when probing the coincidence counts. This effect has strong resemblance to the recently discovered 2p-CBS, despite the photons being generated within the scattering process. To explore the physical origin of this effect and its robustness, we conduct analytical modeling and numerical simulations, systematically varying the distance between the diffuser and the nonlinear crystal. Our results reveal that while the enhancement persists regardless of the generation or scattering order, the width is strongly influenced by the distance the down-converted pairs traverse.

\section{\label{sec:level2} Experiment}
\begin{figure}[hbt!]
 \centering
 \includegraphics[scale=0.5]{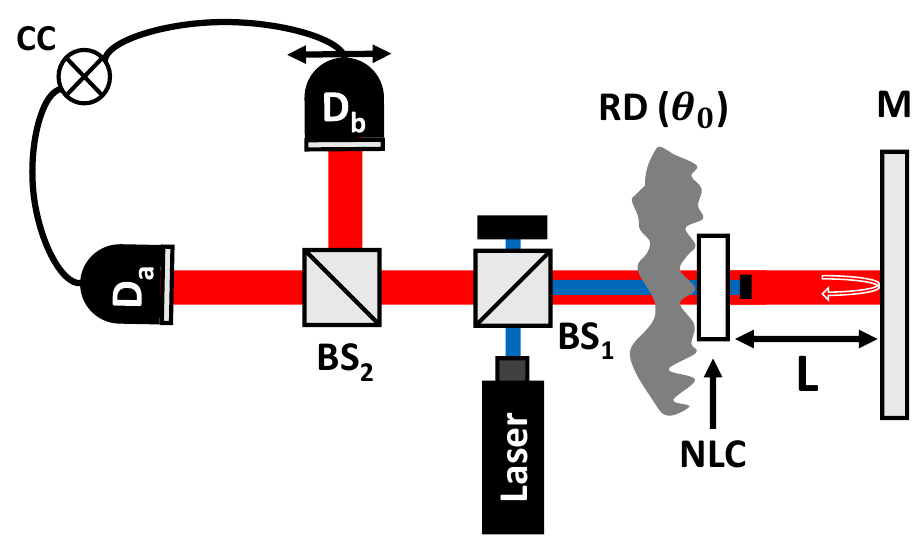}
 \caption{Simplified experimental setup. A pump beam passes through a rotating diffuser (RD), after which a nonlinear PPKTP crystal (NLC) generates entangled photon pairs via SPDC. The pairs propagate a distance $L$ toward a mirror (M), reflect back, and scatter again from the rotating diffuser. Coincidence events are collected using two single-photon detectors: a static detector $\text{D}_a$ and a scanning detector $\text{D}_b$, both positioned in the far-field of the diffuser. BS, beam splitter; $\theta_0$, scattering angle; CC, coincidence circuit.}
 \label{fig:2}
\end{figure}

Figure~\ref{fig:2} depicts a simplified illustration of the experimental setup. A continuous-wave pump beam of waist $W_0\approx500$~$\mu$m and wavelength $\lambda = 405$~nm illuminates a thin rotating diffuser imaged onto a nonlinear crystal. The diffuser chosen here is characterized by a scattering angle of $\theta_0\approx 4.4$~~mrad. This choice of an anisotropic scattering medium allows for an efficient collection of photons using off-the-shelf low numerical aperture lenses, without compromising the underlying physics. Spatially-entangled photon pairs of wavelength $\lambda = 810$~nm are then generated by spontaneous parametric down conversion (SPDC), after which the pump beam is filtered out. The down-converted pairs propagate towards a mirror placed a distance $L$ from the crystal, and get reflected back towards the diffuser. Notice that this configuration (diffuser-mirror) mimics a volumetric scatterer composed of two diffusers at a distance $2L$ \cite{schott2015characterization, cheng2019development,arjmand2021three}. With the inclusion of the nonlinear crystal inside this fictitious volumetric diffuser, it mimics a simplified nonlinear disordered medium. The photon pairs are then measured at the far-field of the diffuser using two single-photon detectors, one static ($D_a$) and the other scanned in the transverse dimension ($D_b$). A coincidence circuit records the number of events in which two photons simultaneously arrive at the two detectors.

\begin{figure*}[hbt!]
 \centering
 \includegraphics[scale=0.5]{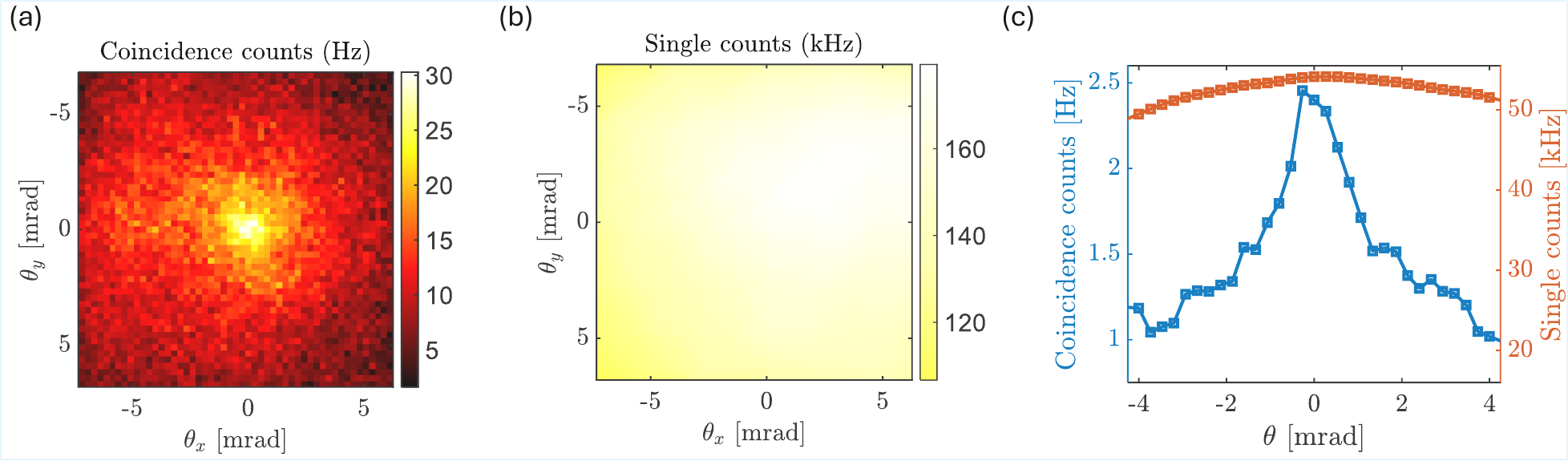}
 \caption{(a) The measured two-dimensional coincidence map at the far-field of the random medium. The coincidence distribution exhibits a clear peak in the backscattering direction. The transverse momenta are expressed in terms of angular positions such that $\theta_{x,y}$ denotes the transverse angular separation between the scanning ($D_b$) and static ($D_a$) detectors. Here, $L=3.5$ ~cm.  (b) The single counts registered by the scanning detector $D_b$ exhibit a homogeneous distribution over the scanned region. Both the coincidence and single counts were acquired simultaneously using 100~$\mu$m core fiber-coupled detectors.  (c) A one-dimensional scan of the peak for the same spacing $L$. Connected blue squares are the resultant coincidence counts, with the connected orange squares being the single counts.  To better resolve the peak, we used 50~$\mu$m core fibers. Here, the data was averaged over two independent scans around the peak. }
 \label{fig:3}
\end{figure*}

Figure~\ref{fig:3}(a) shows the two-dimensional coincidence map as a function of the transverse angular position of the scanning detector $D_b$ ($\theta_b$) relative to the static detector $D_a$ ($\theta_a$), measured at the far-field of the diffuser. Here and throughout the paper, the relative transverse angular separation is denoted by $\theta = \theta_b - \theta_a$. A clear enhanced area of the coincidences, around the origin, is observed, while the single counts distribution, shown in Fig.~\ref{fig:3}(b) reveals no discernible structure and appears nearly homogeneous. The absence of any feature in the single counts, when compared to the coincidence map, is a consequence of entanglement that is distributed across multiple spatial modes\cite{beenakker2009two, peeters2010observation}. To better resolve the peak, we performed one-dimensional scans of the peak with a finer resolution [Fig.~\ref{fig:3}(c)]. Achieving a sufficient signal-to-noise ratio in the coincidence map required addressing two main noise sources: the speckle noise due to scattering and the Poisson noise due to the low flux of photon pairs reaching the detectors. To surpass the latter, we integrated each pixel in the coincidence map for tens of minutes to accumulate thousands of coincidence events per pixel. To surpass the speckle noise, the diffuser was rotated during the acquisition time, achieving a disorder averaged distribution.

\section{\label{sec:level3} Theory}
The above observation raises two questions: (1) Since the effect is characterized by a wider background with a narrower peak on top of it, as well as an almost 2-to-1 enhanced peak relative to the background, how is it related to CBS or its quantum counterpart, the 2p-CBS? (2) Is the observed effect unique to the case where the diffuser is imaged onto the crystal? In what follows, we turn to a theoretical analysis of both questions by studying the above effect as well as more general configurations. Specifically, we consider scenarios in which the thin diffuser and thin crystal are not placed in conjugate planes. To this end, we developed a model and performed numerical simulations to analyze the resulting spatial correlations. Notably, the diffuser–mirror configuration enables us to construct a comprehensive theoretical framework that captures the key processes governing the scattering of entangled photon pairs and the emergence of the enhanced correlation region.

The general two-photon state generated by SPDC in a thin crystal can be described as $\ket{\psi}\propto\sum_{s,i}\Psi_{\text{\textbf{q}}_{s},\text{\textbf{q}}_{i}}\hat{c}_{\text{\textbf{q}}_{s}}^{\dagger}\hat{c}_{\text{\textbf{q}}_{i}}^{\dagger}\ket{0}$, where $\hat{c}_{\text{\textbf{q}}_{i}}^{\dagger}$ is the creation operator of an incident mode with transverse momentum $\textbf{q}_{i}$, and $\Psi_{\text{\textbf{q}}_{s},\text{\textbf{q}}_{i}}$, the two-photon amplitude, is solely determined by the pump and defines the input state \cite{WALBORN201087}. In the special case of a plane-wave pump with zero transverse momentum, the two-photon amplitude reduces to a delta function, $\Psi_{\text{\textbf{q}}_{s},\text{\textbf{q}}_{i}}\propto\delta_{\text{\textbf{q}}_{s},-\text{\textbf{q}}_{i}}$, yielding the well-known Einstein-Podolsky-Rosen (EPR) state \cite{WALBORN201087,einstein1935can}. More generally, however, it is determined by its angular spectrum\cite{WALBORN201087, monken1998transfer}, which is the Fourier transform of the transverse pump field $\psi_p$, impinging on the crystal and evaluated at the sum coordinate, $\Psi_{\text{\textbf{q}}_{s},\text{\textbf{q}}_{i}}=\mathcal{F}\left[\psi_{p}\right]\left(\text{\textbf{q}}_{s}+\text{\textbf{q}}_{i}\right)$. The transverse spatial properties of entangled photon pairs are determined by the coincidence counts which are, in turn, proportional to the two-photon correlation function $\Gamma_{ba} = \overline{\braket{\psi | : \hat{n}_{\textbf{q}_a} \hat{n}_{\textbf{q}_b} : | \psi}}$, where $\hat{n}_{\text{\textbf{q}}}=\hat{d}_{\text{\textbf{q}}}^{\dagger}\hat{d}_{\text{\textbf{q}}}$ is the photon number operator of a reflected mode with transverse momentum $\text{\textbf{q}}$, $:\left(\ldots\right):$ stands for normal ordering, and $\overline{\ldots}$ represents ensemble averaging over different realizations of disorder. 

For the subsequent analysis, it is useful to interpret the correlation function $\Gamma_{ba}$ as the intensity of a two-photon wave function, $\Gamma_{ba}= \vert \Psi^{\text{out}}_{\text{\textbf{q}}_{b},\text{\textbf{q}}_{a}} \vert ^2 $, where $\Psi^{\text{out}}_{\text{\textbf{q}}_{b},\text{\textbf{q}}_{a}}=\bra{0} \hat{d}_{\text{\ensuremath{\boldsymbol{\textbf{q}}_{a}}}} \hat{d}_{\text{\ensuremath{\boldsymbol{\textbf{q}}_{b}}}} \ket{\psi}$. The operators for the outgoing and ingoing modes, $\hat{d}_{\textbf{\ensuremath{\boldsymbol{\text{q}}}}}$ and $\hat{c}_{\textbf{\ensuremath{\boldsymbol{\text{q}}}}}$, are related through classical input-output relations, $\hat{d}_{\textbf{\ensuremath{\boldsymbol{\text{q}}_{a}}}}=\sum_{\text{a}^{\prime}}G_{\textbf{\ensuremath{\boldsymbol{\text{q}}_{a}}},\textbf{\ensuremath{\boldsymbol{\text{q}}_{a^{\prime}}}}}\hat{c}_{\textbf{\ensuremath{\boldsymbol{\text{q}}_{a^{\prime}}}}}$ \cite{RevModPhys.69.731}, where the Green's matrix $G$ describes linear propagation between the crystal and the detectors. We thus obtain the useful form $\Psi^{\text{out}}_{\text{\textbf{q}}_{b},\text{\textbf{q}}_{a}}= \left[G\Psi G^{T}\right]_{_{\boldsymbol{\textbf{q}}_{b},\boldsymbol{\textbf{q}}_{a}}}$.
In the following, we consider two scenarios where the crystal is placed either after or before the scattering layer, and express in both cases the matrices $\Psi$ and $G$ in terms of the transmission matrices representing free-space propagation over a distance $s$ (modeled by a Fresnel kernel) and scattering by the thin diffuser, denoted $H^s(\omega)$ and $V(\omega)$, respectively. 
We explicitly write the dependence on the optical frequency $\omega$ to distinguish between the scattering and propagation of the photon pairs and that of the pump beam.

\begin{figure*}[hbt]
 \centering
 \includegraphics[scale=0.66]{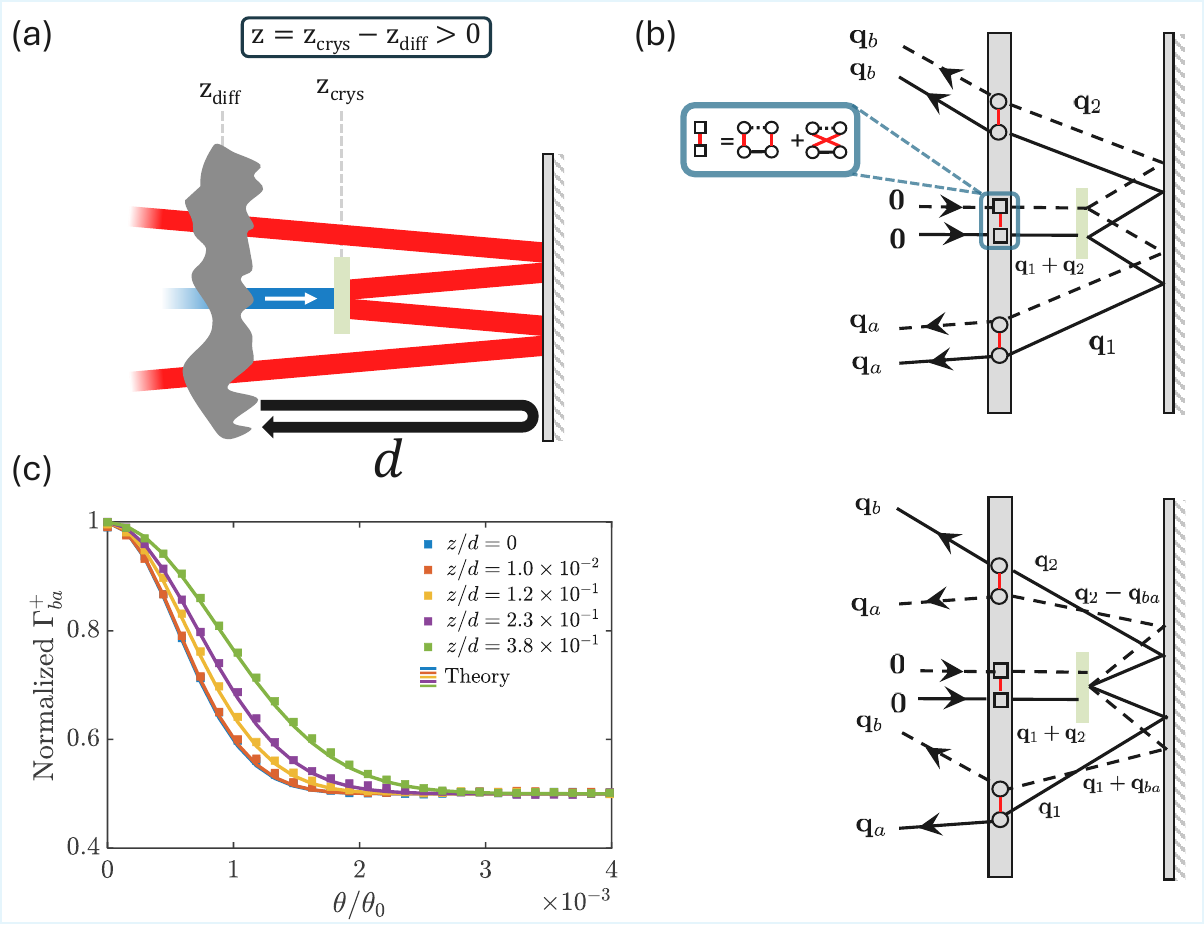}
\caption{(a) Simplified schematic of the experimental setup depicted in Fig.~\ref{fig:2}, with an additional pump-propagation distance $z$ between the diffuser and nonlinear crystal. (b) Diagrammatic representation of the two-leading terms contributing to $\Gamma_{ba}^+$.  Solid (dashed) lines represent the field (conjugate field). Open circles (squares) represent scattering events for the fields at frequency $\omega$ (2$\omega$), while correlated pairs are denoted by a red line connecting them. Each field line or its conjugate propagating carries a weight $H_\omega^z\left ( \textbf{q}\right )$ due to propagating a distance $z$ at frequency $\omega$, while each scattering vertex contributes  $F_\omega\left(\Delta\textbf{q}\right)$ for a momentum transfer $\Delta\textbf{q}$. The weight of each diagram is obtained by assigning transverse momenta to the propagation lines, in accordance with the momentum conservation imposed by the correlators. A summation is then performed over all free transverse momenta. Inset depicts the fact that the correlator at $2\omega$ is a convolution of two correlators at $\omega$. (c) Solid lines: plot of Eq.~(\ref{EqGammaPlusExplicit}), the theoretical prediction for the two-photon correlation function in the $z>0$ case, near the peak. Squares: the numerically obtained values for $\Gamma_{ba}^+$ for the parameters used in the theory. The transverse angular separation is defined as $\theta=\theta_b-\theta_a$. Both the theoretical expression and numerical data were first normalized to unit area, and then by the peak value in the resultant theoretical curve. Moreover, they are shown for $\theta_a=0$. The numerics were realized in the limit where the angular spread exceeds the transverse coherence length. i.e., when $kd\theta_0^2\gg1$. See text for more details.}
  \label{fig:4}
\end{figure*}

\begin{figure*}[hbt]
 \centering
 \includegraphics[scale=0.64]{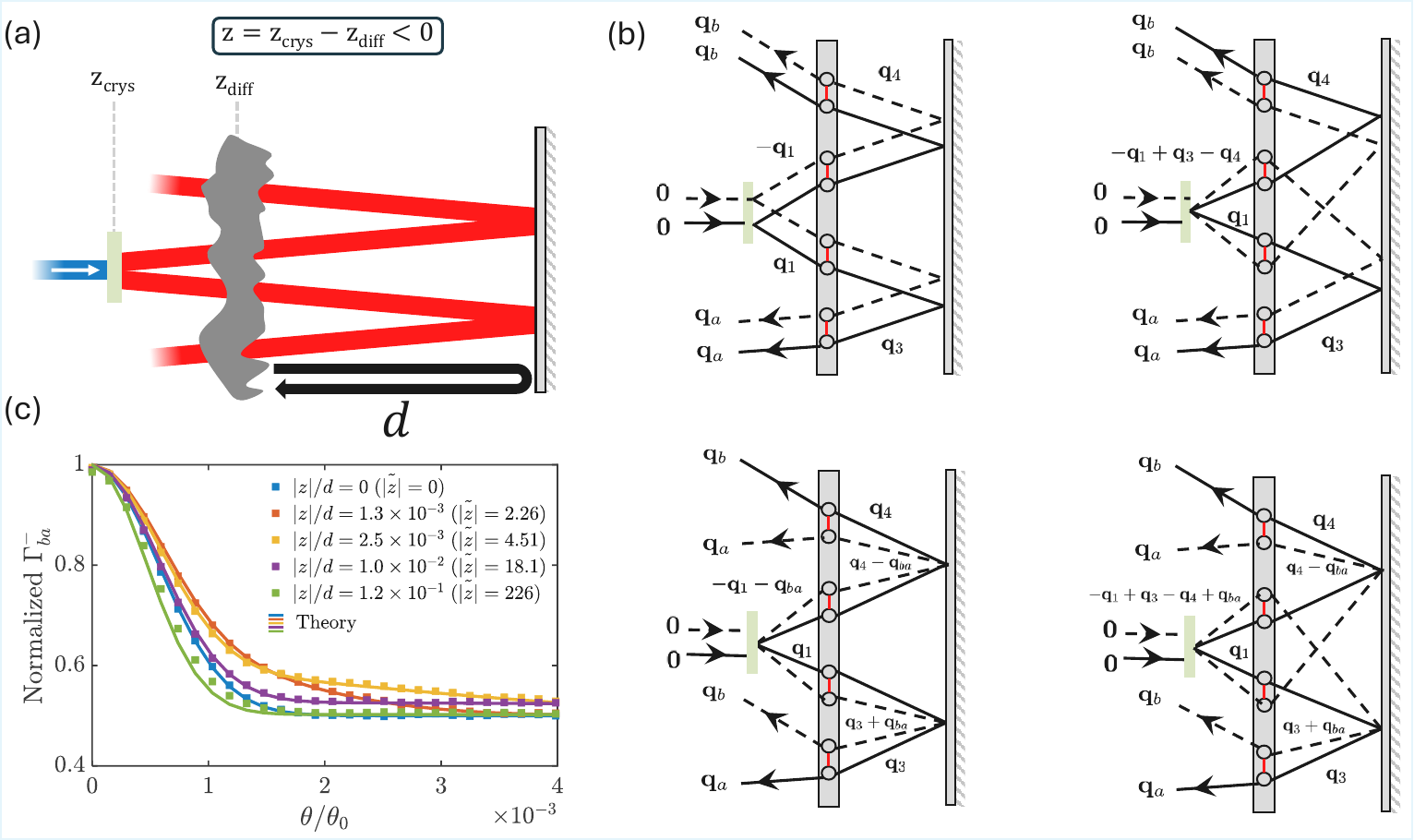}
 \caption{(a) Simplified schematic of the 2p-CBS setup with an additional $|z|$ propagation towards the medium. (b) Diagrammatic representation of the four-leading terms contributing to $\Gamma_{ba}^-$. The diagrammatic rules mentioned in the caption of Fig.~(\ref{fig:4}) can be implemented here as well to arrive at a closed form for $\Gamma_{ba}^-$. (c) Solid lines: Plot of $\Gamma_{ba}^-$, which is a summation of Eq.~(\ref{EqGammaMinusExplicit1}) and Eq.~(\ref{EqGammaMinusExplicit2}). Squares: the numerically obtained values for $\Gamma_{ba}^-$ for the parameters used in the theory. The normalization procedure here is identical to the one mentioned in the caption of Fig.~(\ref{fig:4}). Furthermore, the numerical parameters used here are also identical to the ones shown there.}
 \label{fig:5}
\end{figure*}

In the configuration considered in the experiment  [Fig.~\ref{fig:4}(a)], the Green's matrix takes the form $G^+=V\left(\omega\right)H^{d-z}\left(\omega\right)$ and the two-photon amplitude is given by $\Psi^+_{\ve{q},\ve{q}' }=\left[H^{z}\left(2\omega\right)V\left(2\omega\right)\right]_{\ve{q+q'},\ve{0}}$, where $d=2L$ and $z$ is the axial position of the crystal relative to the diffuser. Here, the crystal is positioned after the diffuser ($z>0$), thereafter denoted as the positive $z$ case. Substitution into the two-photon correlation function yields:
\begin{align} \label{EqGammaPlus}
\Gamma_{ba}^{+}&\propto\overline{\left|\left\{ \left[V\left(\omega\right)H^{d-z}\left(\omega\right)\right]\Psi^{+}\left[V\left(\omega\right)H^{d-z}\left(\omega\right)\right]^{T}\right\} _{\ve{q_b},\ve{q_a}}\right|^{2}}.
\end{align}
By contrast, when the order of the crystal and the diffuser is reversed ($z<0$), a 2p-CBS configuration is obtained, with a propagation distance $\vert z\vert$ between the crystal and the scattering layer [Fig.~\ref{fig:5}(a)]. This case, hereafter, is denoted as the negative $z$ case. The Green's matrix is then given by $G^{-}=V\left(\omega\right)H^{d}\left(\omega\right)V\left(\omega\right)H^{\left|z\right|}\left(\omega\right)$, while for a plane-wave pump the two-photon amplitude is $\Psi^-_{\ve{q},\ve{q}'}\propto\delta_{\ve{q+q'},\ve{0}}$.
Substitution to the two-photon correlation function yields:
\begin{widetext}

\begin{align} \label{eq:435}
\Gamma_{ba}^{-}&\propto\overline{\left|\left\{ \left[V\left(\omega\right)H^{d}\left(\omega\right)V\left(\omega\right)H^{\left|z\right|}\left(\omega\right)\right]\Psi^-\left[V\left(\omega\right)H^{d}\left(\omega\right)V\left(\omega\right)H^{\left|z\right|}\left(\omega\right)\right]^{T}\right\}_{\ve{q_b},\ve{q_a}}\right|^{2}}.
\end{align}
\end{widetext}

To make further progress, we model the thin diffuser as a random phase screen obeying Gaussian statistics. 
Specifically, the real-space transmission matrix elements are written as 
$V_{\boldsymbol{\rho},\boldsymbol{\rho'}}(\omega)=e^{ik D(\boldsymbol{\rho})}\delta_{\boldsymbol{\rho},\boldsymbol{\rho'}}$, 
where $D(\boldsymbol{\rho})$ denotes the diffuser thickness at transverse coordinate $\boldsymbol{\rho}$, and $k=\omega/c$. 
The thickness fluctuations of $D(\boldsymbol{\rho})$ lead to anisotropic scattering within an angular range $\theta_0$. 
We assume that the elements of $V(\omega)$ behave as Gaussian random variables characterized by the correlation function 
$\overline{V_{\boldsymbol{\rho},\boldsymbol{\rho}}(\omega)V_{\boldsymbol{\rho'},\boldsymbol{\rho'}}(\omega)^*}
= e^{-\vert \boldsymbol{\rho}-\boldsymbol{\rho'} \vert^2 / 4\xi_0^2}$, 
where $\xi_0 = 1/k\theta_0$ defines the spatial correlation width of the diffuser. \\
Neglecting material dispersion, we note that 
$V_{\boldsymbol{\rho},\boldsymbol{\rho'}}(2\omega) = V_{\boldsymbol{\rho},\boldsymbol{\rho'}}(\omega)^2$, 
so that the diffuser’s behavior at frequency $2\omega$ follows directly from its properties at $\omega$. 
In particular, 
$\overline{V_{\boldsymbol{\rho},\boldsymbol{\rho}}(2\omega)V_{\boldsymbol{\rho'},\boldsymbol{\rho'}}(2\omega)^*}
= 2\,\overline{V_{\boldsymbol{\rho},\boldsymbol{\rho}}(\omega)V_{\boldsymbol{\rho'},\boldsymbol{\rho'}}(\omega)^*}^2$. 
In the transverse-momentum basis, the correlations of the matrix elements become
\begin{align}
\label{eq:VwV2w}
\overline{V_{\text{\textbf{q}}_{\alpha},\text{\textbf{q}}_{\beta}}\left(\omega\right)
V_{\text{\textbf{q}}_{\gamma},\text{\textbf{q}}_{\delta}}^{*}\left(\omega\right)}
&= F_{\omega}\left(\text{\textbf{q}}_{\alpha}-\text{\textbf{q}}_{\beta}\right)
\delta_{\text{\textbf{q}}_{\beta}-\text{\textbf{q}}_{\delta},\text{\textbf{q}}_{\alpha}-\text{\textbf{q}}_{\gamma}},
\nonumber \\
\overline{V_{\text{\textbf{q}}_{\alpha},\text{\textbf{q}}_{\beta}}\left(2\omega\right)
V_{\text{\textbf{q}}_{\gamma},\text{\textbf{q}}_{\delta}}^{*}\left(2\omega\right)}
&= F_{2\omega}\left(\text{\textbf{q}}_{\alpha}-\text{\textbf{q}}_{\beta}\right)
\delta_{\text{\textbf{q}}_{\beta}-\text{\textbf{q}}_{\delta},\text{\textbf{q}}_{\alpha}-\text{\textbf{q}}_{\gamma}},
\end{align}
with $F_\omega(\ve{q}) \propto e^{-q^2\xi_0^2}$ and 
$F_{2\omega}(\ve{q}) \propto 2 \sum_{\ve{q_\alpha}} F_\omega(\ve{q}-\ve{q}_\alpha) F_\omega(\ve{q}_\alpha)$.  

Within this Gaussian model, both $\Gamma^+_{ba}$ and $\Gamma^-_{ba}$ contain products of four matrices $V(\omega)$ 
and four conjugates $V(\omega)^*$, whose ensemble average reduces, by the complex Gaussian moment theorem, 
to a sum over $4! = 24$ pairwise contractions. 
In what follows, we focus on the regime where, after propagating a distance $d$ the transverse spread of the beam diffracted by the diffuser, $d\theta_0$, 
exceeds the transverse coherence length $\xi_0$. 
Equivalently, the propagation distance $d$ must exceed the longitudinal coherence length 
$z_0 = \xi_0/\theta_0 = 1/k\theta_0^2$. 
In this limit, scattering contributions forcing light to revisit the same coherence area of the diffuser become negligible, 
leaving only four dominant pairwise contractions in $\Gamma^+_{ba}$ and $\Gamma^-_{ba}$, respectively, 
as discussed in detail below. 
\subsection{\label{subsection1} Randomized pump -- $\Gamma^+$}

We first consider the positive-$z$ configuration shown in Fig.~\ref{fig:4}(a). 
In this case, four dominant contractions contribute to the sum in Eq.~\eqref{EqGammaPlus}. 
Since each contraction is doubly degenerate, they can be represented by the two diagrams shown in Fig.~\ref{fig:4}(b). In these diagrams, solid and dashed lines respectively denote the propagating field and its complex conjugate. Open circles (squares) represent scattering events for fields at frequency $\omega$ ($2\omega$), and correlated pairs are connected by red lines. The weight of each diagram is obtained by assigning transverse momenta to the propagation lines, in accordance with the momentum conservation imposed by Eq.~\eqref{eq:VwV2w} (scattering events preserve the momentum difference between a field and its conjugate). Each propagation line carries a weight $H^z_\omega(\ve{q})$, corresponding to free-space propagation over a distance $z$ with transverse momentum $\ve{q}$ and frequency $\omega$, while each scattering vertex contributes a factor $F_{\omega}(\Delta\ve{q})$ for a momentum transfer $\Delta\ve{q}$. A summation is then performed over all free transverse momenta. For instance, the total contribution of the two diagrams in Fig.~\ref{fig:4}(b) reads

\begin{widetext}

\begin{align}
\Gamma^+_{ba} \! \propto \!
    \sum_{\ve{q}_1,\ve{q}_2}\! 
    F_{2\omega}(\ve{q}_2+\ve{q}_1)
    F_{\omega}(\ve{q}_b-\ve{q}_2)
    F_{\omega}(\ve{q}_a-\ve{q}_1)
  \times\left[
    1+
    H^{d-z}_\omega(\ve{q}_2)
    H^{d-z}_\omega(\ve{q}_1+\ve{q}_{ba})^*
    H^{d-z}_\omega(\ve{q}_1)
    H^{d-z}_\omega(\ve{q}_2-\ve{q}_{ba})^*
    \right],
\end{align}
\end{widetext}
where $H^z_\omega(\ve{q}) = e^{iq^2z/2k}$ is the Fresnel kernel associated with the matrix elements 
$H^z_{\ve{q},\ve{q}'}(\omega) = H^z_\omega(\ve{q})\,\delta_{\ve{q},\ve{q}'}$, and $\ve{q}_{ba}=\ve{q}_{b}-\ve{q}_{a}$. 
The degeneracy of each diagram is implicitly contained in the factor $F_{2\omega}(\ve{q}_2+\ve{q}_1)$, 
graphically represented in the inset of Fig.~\ref{fig:4}(b). We also note that the scattering of the pump photons does not introduce any dephasing during propagation between the diffuser and the crystal, since momentum conservation enforces that the field and its complex conjugate propagate in the same direction. Consequently, the dependence of $\Gamma^+_{ba}$ on $z$ arises solely from the dephasing of the photon pair as it propagates over the distance $d - z$ before being scattered by the diffuser into the directions $\textbf{q}_{a}$ and $\textbf{q}_{b}$.

Explicit calculation in three-dimensional space yields
\begin{equation}
\label{EqGammaPlusExplicit}
\Gamma_{ba}^{+} \propto 
2\exp\!\left[-\frac{(\boldsymbol{\theta}_{a}+\boldsymbol{\theta}_{b})^{2}}{4\theta_0^2}\right]
\left\{
1+\exp\!\left[-\frac{(\boldsymbol{\theta}_{a}-\boldsymbol{\theta}_{b})^2}{2\Delta \theta ^+(z)^2}\right]
\right\},
\end{equation}
where the detector positions are specified by their transverse angular positions,
$\boldsymbol{\theta}_{a} = \textbf{q}_{a}/k$ and $\boldsymbol{\theta}_{b} = \textbf{q}_{b}/k$,
and the angular width of the peak is given by: 
\begin{equation}
\Delta \theta ^+(z) = \frac{1}{kd\theta_0(1-z/d)}.
\end{equation}
with $z\leq d/2$. The envelope in Eq.~\eqref{EqGammaPlusExplicit} is centered at
$\boldsymbol{\theta}_{a}+\boldsymbol{\theta}_{b}=0$, since the photon pairs are created in opposite directions with respect to the scattered pump photons, and its width $\theta_0$ reflects the angular range over which both the pump and the photon pairs are scattered.
At the same time, a doubling of the coincidence count occurs for
$\boldsymbol{\theta}_{a}-\boldsymbol{\theta}_{b}=0$, showing that photons tend to emerge from the medium in the same direction, even though they were emitted in opposite directions.    \\
This bunching effect is maintained within an angular range $\Delta \theta ^+(z)$, inversely proportional to the transverse spatial extent $(d-z)\theta_0$ probed by the scattered photons. 
The scaling of this angular range is analogous to that found for the coherent backscattering (CBS) effect, where the angular width is also determined by the inverse of the spatial region explored by the outgoing photons. More precisely and surprisingly, the result in Eq.~\eqref{EqGammaPlusExplicit} is identical to the correlation function obtained for the two-photon coherent backscattering (2p-CBS) configuration studied in Ref.~\cite{safadi2023coherent}, where the crystal is placed immediately before the scattering medium, provided the distance $d$ is replaced by $d-z$.  
This shows that positioning the crystal inside the disordered medium effectively reduces the transverse region probed by the photon pairs, thereby increasing the angular range over which they emerge in the same direction.

 To test the validity of Eq.~(\ref{EqGammaPlusExplicit}), we ran numerical simulations for the two-photon correlation function, as modeled by Eq.~(\ref{EqGammaPlus}). We numerically propagate each photon of the pair through a sequence of diffusers and free-space segments. Free-space propagation is realized by the standard Fresnel propagator. The diffusers are implemented by a complex transmission mask, generated in the Fourier domain by multiplying complex Gaussian random numbers by a Gaussian whose width corresponds to the scattering angle of the diffuser. We note that while the theory assumes a phase-only diffuser, we use this simple model to easily obtain diffusers with a Gaussian distribution in the far-field corresponding to $F_{\omega}(\ve{q})$. The pump is modeled as a wide Gaussian beam, covering thousands of coherence areas of the diffuser, which propagates through the same diffuser and then through free space over a distance $z$. The simulation parameters were chosen to ensure that the angular spread per photon exceeds the transverse coherence length $\xi_0$. To this end, we selected $kd\approx5697$ and $\theta_0\approx0.56$ rad such that $d/z_0\gg1$. Finally, to minimize the effect of speckle noise, the results were averaged over $10^5$  random realizations of the diffuser.
 
 Figure~\ref{fig:4}(c) shows the simulated two-photon corelation $\Gamma_{ba}^{+}$ for various distances $z$ between the crystal and diffuser, together with the corresponding theoretical prediction, obtained by the explicit calculation in Eq.~(\ref{EqGammaPlusExplicit}) for $\boldsymbol{\theta}_a=0$. For each distance, the numerical and theoretical curves were first normalized to unit area and then rescaled by a common factor such that the peak of the theoretical curve equals one. This two-step normalization eliminates arbitrary scaling and reduces noise sensitivity, isolating the lineshape discrepancies. As our theory predicts, we indeed observe a doubling in the coincidences about the origin with a width that is monotonically increasing as $z/d$ grows. The background, however, experiences no significant effect, further validating Eq.~(\ref{EqGammaPlusExplicit}) (a much wider angular range is shown later in Fig.~\ref{fig:7}).

\subsection{\label{subsection2} Plane-wave pump -- $\Gamma^-$}

Next, we study the negative-$z$ configuration, where the crystal is positioned before the diffuser [Fig.~\ref{fig:5}(a)]. In contrast to the positive-$z$ case, the four dominant contributions to $\Gamma_{ba}^{-}$ in Eq.~\eqref{EqGammaPlus} are non-degenerate. They are represented by the four diagrams in Fig.~\ref{fig:5}(b), labeled according to the same transverse momentum conservation rules used in Fig.~\ref{fig:4}(b). This labeling reveals that the two diagrams in the left column (denoted $\Gamma_{ba}^{-(1)}$) behave differently from those in the right column (denoted $\Gamma_{ba}^{-(2)}$). In particular, for $\boldsymbol{\textbf{q}}_{ba} = \boldsymbol{0}$, the photon pairs contributing to $\Gamma_{ba}^{-(1)}$ do not experience any dephasing between the crystal and the diffuser, separated by a distance $\vert z \vert$, whereas such dephasing occurs in both contributions of $\Gamma_{ba}^{-(2)}$. This indicates that $\Gamma_{ba}^{-(2)}$ should vanish once $\vert z \vert$ exceeds the correlation depth $z_0$ of the diffuser. Conversely, when $z = 0$, the two diagrams in each row become indistinguishable (and thus degenerate), leading to the recovery of the 2p-CBS result reported in Ref.~\cite{safadi2023coherent}.

The expression of $\Gamma_{ba}^{-(1)}$ in terms of the correlation functions $F_\omega$ and propagation kernels $H_\omega^z$ can be obtained by applying the same diagrammatic rules as for $\Gamma_{ba}^{+}$. Here, we simply present the final result after explicitly evaluating the summation over all free transverse momenta:
\begin{equation}
\label{EqGammaMinusExplicit1}
\Gamma_{ba}^{-(1)} \propto 
\exp\!\left[-\frac{(\boldsymbol{\theta}_{a}+\boldsymbol{\theta}_{b})^{2}}{4\theta_0^2}\right]
\left\{
1+\exp\!\left[-\frac{(\boldsymbol{\theta}_{a}-\boldsymbol{\theta}_{b})^2}{2\Delta \theta_1^-(z)^2}\right]
\right\},
\end{equation}
where we introduced the angular width
\begin{equation}
\Delta \theta_1^-(z) = \frac{1}{k d \theta_0 (1 + 2\vert z \vert / d)}.
\end{equation}
Similarly to $\Gamma_{ba}^{+}$, $\Gamma_{ba}^{-(1)}$ exhibits an envelope of width $\theta_0$ and photon bunching at $\boldsymbol{\theta}_{a} = \boldsymbol{\theta}_{b}$, which is preserved within an angular range $\Delta \theta_1^-(z)$. We note that $\Delta \theta_1^-(z)$ scales as $\sim 1/(d\theta_0 - 2 z \theta_0)$, whereas $\Delta \theta^+(z) \sim 1/(d\theta_0 - z \theta_0)$. This difference between the negative- and positive-$z$ cases can be understood by comparing the corresponding diagrams: in the negative-$z$ configuration, the photon pair scatters twice from the diffuser, while in the positive-$z$ configuration it scatters only once. Consequently, the deviation of the pair between the crystal and the diffuser, separated by a distance $\vert z\vert$, is typically $2\theta_0$ relative to the output detection angles for $z < 0$, and $\theta_0$ for $z > 0$. This leads to a total transverse spreading of $d\theta_0 - 2 z \theta_0$ for $z < 0$ and $d\theta_0 - z \theta_0$ for $z > 0$.

The evaluation of the two diagrams representing $\Gamma_{ba}^{-(2)}$ yields
\begin{eqnarray}
\label{EqGammaMinusExplicit2}
\Gamma_{ba}^{- (2)} \propto&& \frac{\exp\!\left[-\frac{(\boldsymbol{\theta}_{a}+\boldsymbol{\theta}_{b})^{2} + \tilde{z}^2(3\boldsymbol{\theta}_{a}^2+3\boldsymbol{\theta}_{b}^{2} -2\boldsymbol{\theta}_{a}.\boldsymbol{\theta}_{b})}{4(1+\tilde{z}^2)\theta_0^2}\right]}{\sqrt{1+\tilde{z}^{2}}}\nonumber\\
\times&&\left\{
1+\exp\!\left[-\frac{(\boldsymbol{\theta}_{a}-\boldsymbol{\theta}_{b})^2}{2\Delta \theta_2 ^-(z)^2}\right]
\right\},
\end{eqnarray}
where $\tilde{z}= z/z_0$ denotes the position of the crystal in units of the correlation depth of the diffuser, and the angular width is
\begin{equation}
\Delta \theta_2 ^-(z) = \frac{\sqrt{1+\tilde{z}^{2}}}{kd\theta_0\sqrt{1+2\vert z \vert/d}}.
\end{equation}

As anticipated, $\Gamma_{ba}^{-(2)}$ carries a global weight $\propto 1/\sqrt{1+\tilde{z}^2}$, which vanishes for $z \gg z_0$. 
In addition, unlike $\Delta \theta_1^-(z)$, the width $\Delta \theta_2^-(z)$ depends not only on $z/d$ but also on $z/z_0$. 
In particular, for $\vert z \vert \lesssim z_0 \ll d$, $\Delta \theta_2^-(z)$ increases with $\vert z \vert$. 
This reflects the fact that, in the bottom-right diagram of Fig.~\ref{fig:5}(b), the entangled photons generated by the crystal are initially constrained to traverse the same coherence area of size $\xi_0$ within the diffuser. 
As a result, the effective transverse region probed by the outgoing photons is reduced, which broadens the angular range of photon bunching. 
Since $\Delta \theta_1^-(z)$ remains nearly constant for $\vert z \vert \lesssim z_0$, we thus expect the overall width of $\Gamma_{ba}^-$ to increase in this regime due to the growth of $\Delta \theta_2^-(z)$. 
Conversely, for $\vert z \vert \gg z_0$, the contribution of $\Gamma_{ba}^{-(2)}$ becomes negligible, and the width of $\Gamma_{ba}^-$ decreases following the behavior of $\Delta \theta_1^-(z)$. 

Figure~\ref{fig:5}(c) shows results of numerical simulations  for the two-photon correlation $\Gamma_{ba}^-$ [Eq.~(\ref{eq:435})], exhibiting the non-monotonic behavior discussed above. Apart from the reversal of scattering order and propagation paths, the normalization procedure and parameters are identical to those used for $\Gamma_{ba}^+$. Solid curves are the explicit calculation of $\Gamma_{ba}^-=\Gamma_{ba}^{-(1)}+\Gamma_{ba}^{-(2)}$ [sum of Eq.~(\ref{EqGammaMinusExplicit1}) and Eq.~(\ref{EqGammaMinusExplicit2})]. As expected, a sharp doubly-enhanced peak is observed. The peak for $z=0$ (blue) is identical to peak of $\Gamma_{ba}^+$ in Fig.~\ref{fig:4}(c) (blue). Beyond $z=0$, however, the width of $\Gamma_{ba}^-$ broadens when $\tilde{z}\sim1$ due to the contribution of $\Delta\theta_2^-(z)$ mentioned above, and tightens again for $|\tilde{z}|\gg1$ due to $\Delta\theta^-_1(z)$ and the fact that the relative weight of $\Gamma_{ba}^{-(2)}$ decreases with $|\tilde{z}|$. 
\begin{figure}[hbt!]
 \centering
 \includegraphics[scale=0.5]{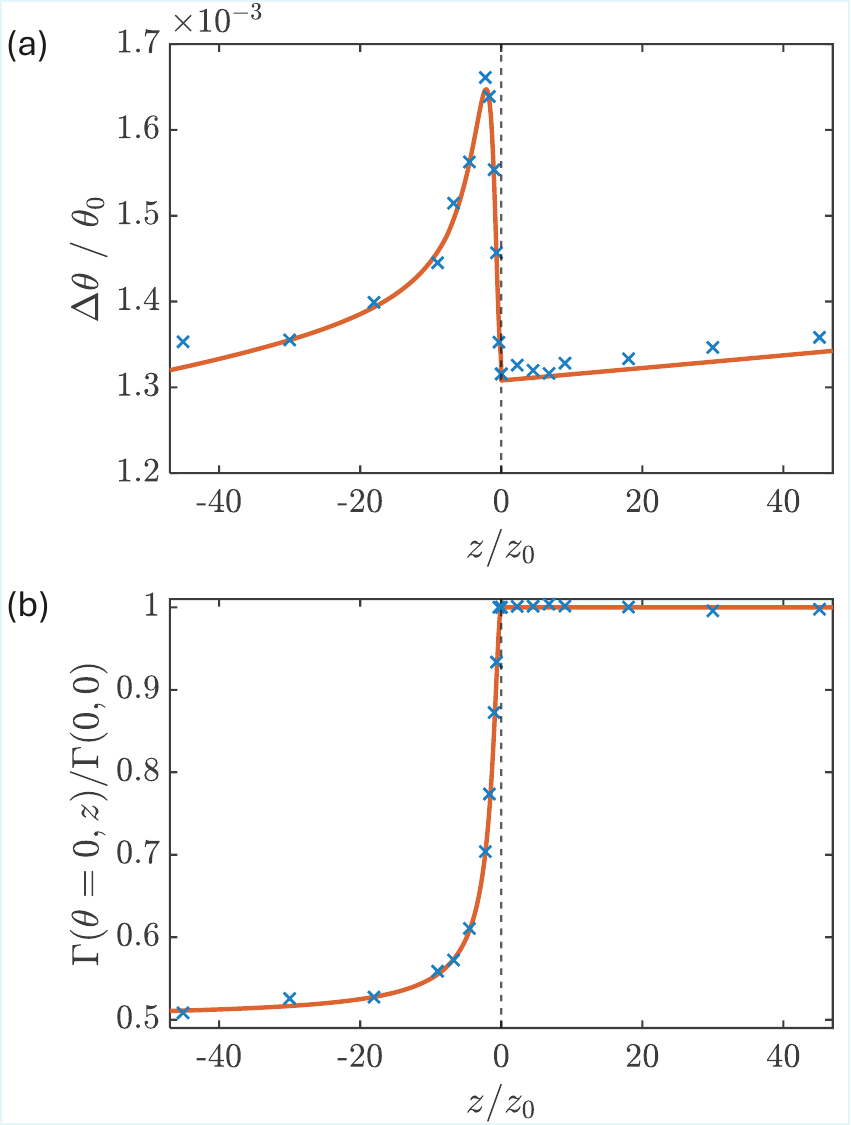}
 \caption{(a) The width, $\Delta\theta$, defined as the full-width-at-half-maximum (FWHM) of the enhancement terms in $\Gamma_{ba}^\pm$, is extracted from numerical simulations (blue crosses) and compared to the theoretical prediction (orange line): for $\Gamma_{ba}^+$ the theory is given by Eq.~(\ref{EqGammaPlusExplicit}), and for $\Gamma_{ba}^-$ by the sum of Eqs.~(\ref{EqGammaMinusExplicit1}) and (\ref{EqGammaMinusExplicit2}). To isolate the enhancement peak, we subtracted from the obtained two-photon correlation $\Gamma_{ba}^\pm$, the background terms in the aforementioned equations. The FWHM was normalized by $\theta_0.$ (b) The amplitude of $\Gamma_{ba}^\pm$ at $\theta=0$. The theory (orange line) as well as the simulation  (blue crosses) were both normalized by the peak value at $z=0$ separately.}
 \label{fig:5p5}
\end{figure}
\subsection{\label{subsection3} Comparison}
To compare $\Gamma_{ba}^+$ and $\Gamma_{ba}^-$, we consider two figures of merit. First, Fig.~\ref{fig:5p5}(a) shows the full width at half maximum (FWHM) of the peak in the two-photon correlation function, $\Delta \theta$, expressed in units of $\theta_0$. The peak profile was obtained by subtracting the corresponding theoretical background from the simulation data (blue crosses). For reference, we also plot the theoretical width, derived in the same manner (orange solid line). As discussed above, while for $\tilde{z}>0$ the width increase monotonically across the entire range, for $\tilde{z}<0$  it exhibits a non-monotonic behavior, with a peak at $\vert \tilde{z}\vert \approx 2.1$. Second, in Fig.~\ref{fig:5p5}(b) we show the peak coincidences normalized by the peak value of $\Gamma_{ba}^\pm$ at $z=0$ (blue crosses). Once again, the theoretical amplitude is shown for reference (orange curve). As anticipated, the enhancement of $\Gamma_{ba}^+$ is completely independent of $\tilde{z}$, while it experiences a sharp drop for increasing values of $|\tilde{z}|$ for $\Gamma_{ba}^-$, after which it plateaus at about 0.5. This can be easily understood from both contributions to $\Gamma_{ba}^-$, as the amplitude of $\Gamma_{ba}^{-(2)}$ vanishes for $|\tilde{z}|\gg1$ while that of $\Gamma_{ba}^{-(1)}$ is constant contributing half the total amplitude.
\begin{figure}[hbt!]
 \centering
 \includegraphics[scale=0.54]{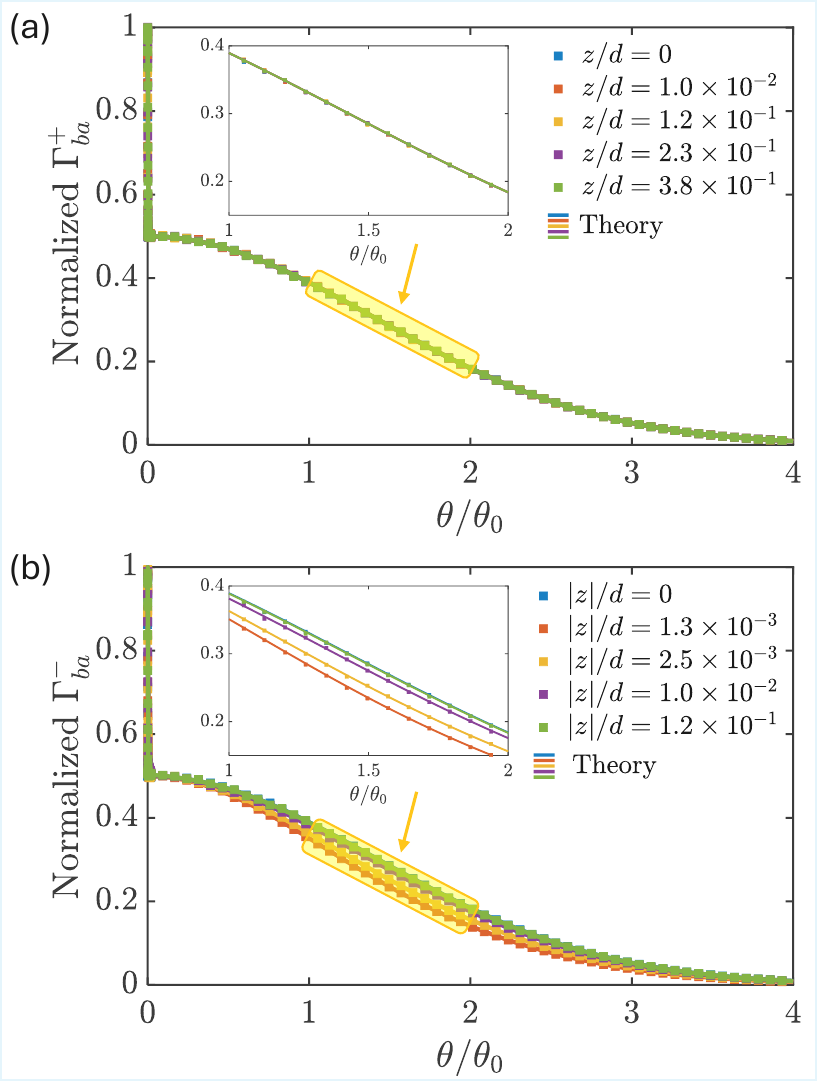}
 \caption{The two-photon correlation $\Gamma^{+}_{ba}$ (a) and $\Gamma^{-}_{ba}$ (b), presented over a wider range than in Fig.~(\ref{fig:4}) and Fig.~(\ref{fig:5}). 
 Insets show a magnified view of the shaded yellow region. For clarity, the marker density of the background was diluted to probe the behavior more clearly.}
 \label{fig:7}
\end{figure}

Up to this point, we have primarily focused on the behavior of the peak of $\Gamma_{ba}^\pm$. For a complete characterization of the two-photon correlation functions, it is also important to assess the impact on the broad background. Figure (\ref{fig:7}) shows a much wider angular range than presented thus far, exhibiting the wide background of $\Gamma_{ba}^+$ and $\Gamma_{ba}^-$. Insets display a zoomed-in portion of a background section (yellow shaded area), far from the peak. While the background of $\Gamma_{ba}^+$ remains entirely unaltered for $z\neq0$, in agreement with Eq.~(\ref{EqGammaPlusExplicit}), $\Gamma_{ba}^-$ undergoes a significant transition. Specifically, the background tightens in the regime $\vert z \vert \lesssim z_0$ regime [blue to orange data in Fig.~\ref{fig:7}(b)], and as $\vert \tilde{z} \vert$ increases further (orange to green data) it relaxes back to the form observed at $z=0$ (blue data), where the  background width is set by $\theta_0$. These features are consistent with the first terms in Eq.~\eqref{EqGammaMinusExplicit1} and~\eqref{EqGammaMinusExplicit2}, which predict that the envelope of $\Gamma_{ba}^{-(1)}$ has a constant width $\theta_0$ independent of $z$, whereas the envelope of $\Gamma_{ba}^{-(2)}$ has a width $\sim \sqrt{(1+\tilde{z}^2)/(1+3\tilde{z}^2)}\theta_0$ and a weight $\sim 1/\sqrt{1+\tilde{z}^2}$. 

\section{Conclusions and outlook}
In conclusion, we experimentally and theoretically studied pair generation within a dynamically varying volumetric scatterer. We observed that, in this scenario, the pairs exhibit an enhancement of their two-photon correlations. We then generalized our findings to configurations where the diffuser and nonlinear crystal are placed at arbitrary positions relative to each other, covering cases such as pump scattering followed by pair generation and scattering, or unaltered pair generation followed by double scattering of the pairs. This generalization reveals the rich underlying physics of these processes, as supported by our numerical simulations. Even in this simplified engineered model of the medium, the observed correlations are sensitive to the sequence of scattering and pair generation, highlighting the complexity of their interplay inside thick nonlinear scattering medium. We anticipate that this first step toward modeling photon-pair generation in the presence of multiple scattering, using a diagrammatic approach, will prove useful to uncover new phenomena and enable efficient pair generation within complex media.

\section*{Funding}
This research was supported by the Zuckerman STEM Leadership Program, the Israel Science Foundation (grant No. 2497/21), and by the program ``Investissements d’Avenir'' launched by the French Government. M.S acknowledges the support of the Israeli Council for Higher Education (VATAT).

\bibliography{refs}

\end{document}